\documentclass[12pt]{article}
\usepackage{latexsym}
\usepackage{amsmath,,calrsfs}
\usepackage{amsfonts}
\usepackage{amssymb}
\usepackage{amscd}
\usepackage{bbm}
\usepackage{fancybox}
\usepackage{cite}
\usepackage{amsmath,amsfonts,amsbsy}
\usepackage{pstricks,pst-node}
\usepackage[small,bf,hang]{caption2}
\usepackage{graphicx}
\usepackage{epsfig}
\usepackage{psfrag}
\usepackage{comment}

\usepackage{float}

\psset{unit=1.3cm,linewidth=.5pt,radius=.2}  

\usepackage{multirow}                     
\usepackage{float}                          
\usepackage{lscape}                         
\usepackage{bm}

\newcommand{\bb}{\bar\beta}

\newcommand{\beq}{\begin{equation}}
\newcommand{\eeq}{\end{equation}}
\newcommand{\bi}{\begin{itemize}}
\newcommand{\ei}{\end{itemize}}
\newcommand{\bt}{\begin{tabular}}
\newcommand{\et}{\end{tabular}}
\newcommand{\bc}{\begin{center}}
\newcommand{\ec}{\end{center}}

\newcommand{\be}{\begin{equation}}
\newcommand{\ee}{\end{equation}}
\newcommand{\bea}{\begin{eqnarray}}
\newcommand{\eea}{\end{eqnarray}}
\newcommand{\ba}{\begin{array}}
\newcommand{\ea}{\end{array}}

\def\bbox{{\,\lower0.9pt\vbox{\hrule \hbox{\vrule height 0.2 cm
\hskip 0.2 cm \vrule height 0.2 cm}\hrule}\,}}
\newcommand{\dsl}{\pa \kern-0.5em /}

\font\mybb=msbm10 at 10pt
\def\bb#1{\hbox{\mybb#1}}
\def\bZ {\bb{Z}}

\def\bH {\bb{H}}

\def\bX {\bb{X}}
\def\bP {\bb{P}}

\makeatletter \@addtoreset{equation}{section} \makeatother

\def\slashchar#1{\setbox0=\hbox{$#1$}           
   \dimen0=\wd0                                 
   \setbox1=\hbox{/} \dimen1=\wd1               
   \ifdim\dimen0>\dimen1                        
      \rlap{\hbox to \dimen0{\hfil/\hfil}}      
      #1                                        
   \else                                        
      \rlap{\hbox to \dimen1{\hfil$#1$\hfil}}   
      /                                         
   \fi}

\begin{document}

\begin{titlepage}
\begin{center}

\hfill  UMTG-32, DAMTP-2014-2

\vskip 1.5cm

{\Large \bf  All Superparticles are BPS}

\vskip 1cm

{\bf Luca Mezincescu\,${}^1$, Alasdair J. Routh\,${}^2$ and 
Paul K.~Townsend\,${}^2$} \\

\vskip 25pt

{\em $^1$ \hskip -.1truecm
\em Department of Physics
University of Miami, \\ Coral Gables, FL 33124, USA\vskip 5pt }
DAMTP-2014-2
{email: {\tt L.Mezincescu@server.physics.miami.edu}} \\

\vskip .4truecm

{\em $^2$ \hskip -.1truecm
\em  Department of Applied Mathematics and Theoretical Physics,\\ Centre for Mathematical Sciences, University of Cambridge,\\
Wilberforce Road, Cambridge, CB3 0WA, U.K.\vskip 5pt }

{email: {\tt A.J.Routh@damtp.cam.ac.uk, P.K.Townsend@damtp.cam.ac.uk}} \\

\end{center}

\vskip 0.5cm

\begin{center} {\bf ABSTRACT}\\[3ex]
\end{center}

The generic action for an $N$-extended superparticle  in $D$-dimensional Minkowski spacetime  is shown to have
``hidden'' supersymmetries (related  by ``dualities''  to the manifest supersymmetries) such that the full supersymmetry algebra is BPS-saturated;
the exceptions (which include, trivially, the massless case) are those  for which the manifest supersymmetry algebra is already BPS-saturated. 
Moreover,  it is shown that any ``non-BPS''  superparticle action  is a gauge-fixed version of the ``BPS''  superparticle action for which all supersymmetries
are manifest.  An example is the $N=1$ massive $D=10$ superparticle, which actually has $N=2$ supersymmetry and is equivalent to the action for a  D0-brane of IIA  superstring theory.

\end{titlepage}

\newpage
\setcounter{page}{1} 

\newpage

\section{Introduction}

The superparticle action describes, classically, the motion of a  particle in some superspace extension of a $D$-dimensional spacetime. In the case of a 
Minkowski vacuum, the action has a  manifest super-Poincar\'e invariance, and quantization yields a  supermultiplet of particle states. 
The first such action to be considered was that of the massless $D=10$ superparticle \cite{Brink:1981nb}, and it was  found that it has a non-manifest ``fermionic''  gauge  
invariance \cite{Siegel:1983hh}, now usually  called ``kappa-symmetry''.  For $N=1$ supersymmetry, the massless $D=10$ superparticle yields, upon quantization,  the Maxwell 
supermultiplet. For $N={\rm IIA}$ or $N={\rm IIB}$ supersymmetry it yields the corresponding $D=10$ graviton supermultiplet. 

Here we will be concerned, principally, with the massive superparticle. In some respects this is simpler  than the massless superparticle because, generically, massive superparticle actions do not have fermionic gauge invariances. However, as we shall see, there is a close connection between the massless and massive cases.  The results we present are valid for a general spacetime dimension but because properties of spinors are dimension dependent it is not convenient to consider all dimensions at once.  For simplicity of presentation, we will assume that $D=3,4$ or  $10$ (mod $8$)  in which case we may assume that the spinor coordinate $\Theta$ of $N=1$ superspace is  Majorana, with a Majorana conjugate, $\bar\Theta = \Theta^T C$, where the charge conjugation matrix $C$ is antisymmetric, and also that $\Theta$ is chiral or anti-chiral if $D=10$.  With this understood\footnote{Later we shall consider the case of $D=6$  to illustrate the differences that arise when the minimal spinor is not Majorana or Majorana-Weyl.}, the action, in Hamiltonian form, for the $N=1$ superparticle of mass $m$ is
\be\label{N=1}
S= \int\! dt \left\{ \dot X\cdot P  + i \bar\Theta\slashchar{P} \dot\Theta  - \frac{1}{2} e\left(P^2+m^2\right)\right\}\, ,
\ee
where $\Gamma$ are the Dirac matrices, and an overdot indicates a derivative with respect to the arbitrary worldline coordinate $t$. The ``einbein'' $e$ is a Lagrange multiplier for 
the mass-shell constraint.   The orthosymplectic two-form defined by this action is
\be\label{Om1}
\Omega= d\left[P\cdot \Pi\right] = dP\cdot \Pi + id\bar\Theta \slashchar{P} d\Theta\, ,
\ee
where 
\be
\Pi=   dX + i\bar\Theta\Gamma d\Theta\, , 
\ee
and the exterior product of forms is implicit.  The 1-forms $\Pi$ and $d\Theta$ are invariant under the infinitesimal supertranslations 
\be\label{manifestsusy}
\delta X = a+ i\bar{\Theta}\Gamma^{\mu}\epsilon \,, \qquad \delta \Theta = \epsilon \, , 
\ee
where $a$ is a constant $D$-vector parameter and $\epsilon$ a constant anticommuting Majorana spinor parameter. This makes manifest the super-Poincar\'e invariance of 
$\Omega$, since $P$ is super-Poincar\'e inert.
The Noether charges that span the supertranslation subalgebra are
\be
\mathcal{P} = P\, , \qquad \mathcal{Q} = \slashchar{P}\Theta\,.
\ee
Provided that $m\ne0$, the two-form $\Omega$ is invertible on the mass shell and we can invert it to find the Poisson brackets (PBs) of the canonical variables. These can be used to show (for $D=3,4$ mod $8$) that 
\be\label{N=1alg}
\left\{ \mathcal{Q}_\alpha,\mathcal{Q}_\beta\right\}_{PB} = -i\left(C\Gamma \right)_{\alpha\beta} \cdot \mathcal{P}\, . 
\ee
In the $D=10$ case, ${\cal Q}$ is chiral, with a chirality opposite to that of $\Theta$,  so the right-hand side of (\ref{N=1alg})  is modified to include a chiral projection matrix.  Apart from this, 
the details are essentially the same as those spelled out for $D=3$ in \cite{Mezincescu:2010yq} , where it was shown that quantization yields  the $N=1$ semion supermultiplet of  helicities $(-\frac{1}{4},\frac{1}{4})$.

In this paper we make two observations about massive superparticle actions, such as (\ref{N=1}). The first is that they  have additional ``hidden'' (i.e. non-manifest) 
supersymmetries, such that the full supersymmetry algebra is the same as that of a higher-dimensional massless superparticle that has been dimensionally reduced by 
fixing the components of the momentum in the extra dimension(s), which then appear in the full lower-dimensional supersymmetry algebra as central charges. Unitarity implies
a bound on the mass  in terms of the central charges, which we shall refer to as the BPS bound\footnote{The bounds originally found by  Bogomolnyi  and  Prasad and Sommerfield   were  bounds on the energy of a supersymmetrizable field theory in terms of topological charges, which were later shown by Witten and Olive to appear in the supersymetry algebra  of the supersymmetrized field theory \cite{W&O}.  The BPS bounds are then also implied by the supersymmetry algebra, as is the case here.} and the  construction described leads to a supersymmetry algebra in which this bound is saturated. In this sense, all massive superparticles ``are BPS'', as are (trivially) massless superparticles, hence our  title. 

The hidden supersymmetries of massive superparticle actions are related to the manifest ones by a ``duality''.  In the $D=3,4$ cases this is a self-duality in the sense that the  massive superparticle action is mapped into itself; this implies a $\bZ_2$ automorphism of the full supersymmetry algebra. In those $D=10$ cases in which the action is  chiral the duality flips the chirality, thereby showing that the chiral and anti-chiral actions are equivalent. In particular, the $D=10$ superparticle actions with $(1,0)$ and $(0,1)$ supersymmetry are equivalent, and both actually have $(1,1)$ supersymmetry; in this case the full supersymmetry algebra is exactly that of the $N={\rm IIA}$ superparticle, i.e. D0-brane, which indeed
has an action that is the dimensional reduction of the action for a massless 11-dimensional superparticle \cite{Bergshoeff:1996tu}, and this action realizes the $N={\rm IIA}$ supersymmetry {\it manifestly} but at the cost of having unphysical fermionic variables that can be ``gauged away'' by a kappa-symmetry gauge transformation \cite{deAzcarraga:1987dh}.

As this D0-brane example illustrates, there are at least two massive superparticle actions with exactly the same full supersymmetry algebra. One is the  massive superparticle
with hidden supersymmetries, but no gauge invariances. The other realizes the full supersymmetry algebra manifestly but has a fermionic gauge invariance.
In the latter case the central charge of the supersymmetry algebra appears as the result of a ``Wess-Zumino mass term'' \cite{Azcarraga:1982dw}. For the {\it generic} 
$N=2$ superparticle this term has an arbitrary coefficient $q$, and the action is 
\be\label{N=2alt}
S= \int \! dt \left\{ \dot X\cdot P + i \bar\Theta^a \left(\slashchar{P} \delta^{ab} + q\varepsilon^{ab}\right) \dot\Theta^b - \frac{1}{2} e\left(P^2+m^2\right) \right\}\, , 
\ee
where the index $a=1,2$ is summed over.  This has an $N=2$ super-Poincar\'e invariance, and the two spinor Noether charges  are 
\be
\mathcal{Q}^a = \slashchar{P}\Theta^a + q \varepsilon^{ab}\Theta^b \,  \qquad (a=1,2).
\ee
The orthosymplectic form defined by the $N=2$ action is 
\be\label{Om2}
\Omega= d\left[ P \cdot\Pi + iq\varepsilon^{ab} \bar\Theta^a d\Theta^b \right] =
dP \cdot\Pi + i d\bar\Theta^a \left(\slashchar{P}\delta^{ab} +q\varepsilon^{ab}\right) d\Theta^b \, , 
\ee
which is manifestly $N=2$ super-Poincar\'e invariant.  It  is invertible provided that $m\ne|q|$, and its inverse yields the Poisson brackets of the canonical variables. The  
Poisson bracket of $P$ with $\Theta$ is zero, for example, and 
\be\label{thetaPBs}
\left\{\Theta^{a \, \alpha} , \Theta^{b\, \beta}\right \}_{PB}= \frac{1}{\left(m^2-q^2\right)} \left[\left( \slashchar{P} \delta^{ab} - q\varepsilon^{ab}\right)C\right]^{\alpha\beta}\, . 
\ee
Using these relations, one  finds that 
\be\label{N=2susy}
\left\{ \mathcal{Q}^a_\alpha,\mathcal{Q}^b_\beta \right\}_{PB} = -i\left[C\left(\slashchar{P} \delta^{ab}+ q \varepsilon^{ab} \right)\right]_{\alpha\beta}\, . 
\ee
This is for $D=3,4$ mod $8$ because an additional chirality projection matrix is needed on the right hand side for $D=10$.  Choosing $C=\Gamma^0$  we find, in the rest frame, after passing to the quantum theory that the  matrix of anticommutators of the supersymmetry charges is   $m\delta^{ab} + q\Gamma^0\varepsilon^{ab}$,  which is positive semi-definite only if $m\ge |q|$, which is the BPS bound for this $N=2$ supersymmetry algebra.   

When $m=|q|$ the action (\ref{N=2alt})  is no longer in Hamiltonian form because $\Omega$ ceases to be invertible. This is related to the fact that the BPS-saturated massive $N=2$ superparticle action has a kappa-symmetry gauge invariance, as could be anticipated from  its interpretation as a dimensionally reduced massless superparticle. Because of this gauge invariance, the BPS-saturated $N=2$ superparticle has a physical phase space of the same (graded)  dimension as the $N=1$ superparticle.  In fact, twistor methods have been used to show  for both $D=3$ \cite{Gorbunov:1996ed,Mezincescu:2013nta} and $D=4$ \cite{Mezincescu:2013nta} that the two massive superparticle actions are equivalent. 

The second observation of this paper is that this equivalence holds not just for $D=3,4$,  but for {\it any} dimension.  The $N=1$ massive superparticle is just a gauge-fixed version of the 
$N=2$ BPS superparticle, which explains why the former has ``hidden'' supersymmetries. This by itself is not sufficient for equivalence because a gauge fixing that breaks manifest  Lorentz invariance would need to be followed by a field redefinition that restores it, and then there is no guarantee that the Lorentz Noether charges of the $N=2$ BPS superparticle will coincide with those of the massive $N=1$ superparticle. However, it is possible to fix the kappa-symmetry  Lorentz-covariantly; this is not possible for a massless superparticle but it becomes
possible upon dimensional reduction, i.e. for a BPS superparticle,  because the relevant Lorentz group is then that of the lower dimension\footnote{As far as we are aware, the possibility of a Lorentz-covariant gauge for massive BPS superparticles has been noted previously only for the $D=2$ case \cite{AGIT}.}. When the kappa-symmetry is gauge fixed in this way, 
the $N=2$ BPS action becomes equivalent to the massive $N=1$ action.

More generally, there is a multiple equivalence of massive superparticle actions, all with the same (graded) dimension of their physical phase space and all with the same
BPS-saturated full supersymmetry algebra. Starting with the massless superparticle in any dimension we can dimensionally reduce to get a BPS-saturated massive superparticle in a lower dimension, with a kappa-symmetry that allows half of the fermionic variables to be ``gauged away''.  If this has $N=4$ supersymmetry, for example, we can now gauge-fix
the kappa-symmetry to get an action for which only $N=2$ supersymmetry is manifest; this is equivalent to the obvious $N=2$ extension of (\ref{N=1}), i.e. the $q=0$ case of 
(\ref{N=2alt}). However, we could also partially gauge fix to get an action for which $N=3$ supersymmetry is manifest, leaving a residual kappa-symmetry. This is equivalent to the 
maximally BPS-saturated $N=3$ superparticle action, which actually has a hidden $N=4$ supersymmetry.  For $N>4$ there are many more possibilities.

\section{Hidden supersymmetries from dualities} 

The $N=1$ massive superparticle action  (\ref{N=1}) is ``self-dual''  in the sense that it is mapped into itself by the transformation
\be\label{dual}
\Theta \mapsto m^{-1}\slashchar{P}\Theta \,, \qquad e \mapsto e - 2im^{-2}\bar\Theta\slashchar{P}\dot\Theta \,.
\ee
Strictly speaking, this is not a duality transformation because the dual of the dual of $\Theta$ is $-\Theta$ (on the mass shell)   but we get a duality in the strict sense by combining the above transformation with a rotation by $\pi$ since a $2\pi$ rotation changes the sign of $\Theta$. However, it is simpler  to work with the ``duality'' transformation as  given. Given the mass-shell constraint $P^2=-m^2$, the supersymmetry Noether charge $\mathcal{Q}$ is mapped to the new spinor charge 
\be
\tilde{\mathcal{Q}} = -m\Theta \,. 
\ee
With respect to  the Poisson brackets found by inverting the 2-form $\Omega$ of  (\ref{Om1}), one can show that this new spinor charge generates the following 
infinitesimal transformations of the canonical variables
\be\label{non-manifest}
\delta_{\eta} X^{\mu} = im^{-1}\bar{\eta}\Gamma^{\mu}\slashchar{P}\Theta \,, \qquad \delta_{\eta} \Theta = m^{-1}\slashchar{P}\eta\,, 
\ee
where $\eta$ is another constant anticommuting Majorana spinor parameter.  The action is invariant if these transformations are supplemented by the following  transformation of the einbein:
\be
\delta_\eta e = - 2im^{-1} \bar \eta \dot\Theta\, . 
\ee
The $N=1$ massive superparticle action therefore has  $N=2$ supersymmetry\footnote{It might appear that we could rescale the parameter $\eta$ of (\ref{non-manifest})  before taking $m\to0$ to find an  additional supersymmetry of the massless superparticle action (and these certainly exist as ``off-shell'' symmetries \cite{Brink:1981rt}) but this rescaling also rescales the ``hidden'' Noether charge to  $-m^2\theta$,  which is zero for $m=0$. The massless superparticle has no ``on-shell''  hidden supersymmetries.}. Moreover,  by a computation of the Poisson brackets of the components of both spinor charges, 
$\mathcal{Q}$ and $\tilde{\mathcal{Q}}$, one can show that the $N=2$ supersymmetry algebra is that of (\ref{N=2susy}) with $m=|q|$,  so the BPS bound is saturated.
We conclude that the massive $N=1$ superparticle has a hidden $N=2$ supersymmetry, with a BPS-saturated $N=2$ supersymmetry algebra. 

In the $D=10$ case, where $\Theta$ is Majorana-Weyl, the transformation (\ref{dual}) changes the chirality of $\Theta$ and hence maps the massive superparticle action with $(1,0)$ supersymmetry into the one with $(0,1)$ supersymmetry. This shows that these two $N=1$ massive superparticle actions  are equivalent. In addition, the Noether charge
corresponding to the manifest supersymmetry of one action is mapped to the Noether charge corresponding to the hidden supersymmetry of the other action, so either action actually has two supersymmetry charges, of opposite chirality.  Thus, it is still true that the $N=1$ superparticle actually has $N=2$ supersymmetry, but this is a non-chiral 
$(1,1)$ supersymmetry. 

The duality of the $N=1$ superparticle action can be extended to the generic $N=2$ superparticle action (\ref{N=2alt}). Provided that $m\ne|q|$, this action is mapped to itself under the duality transformation
\be\label{dual2}
\Theta^1 \mapsto \frac{\slashchar{P}\Theta^2 -q\Theta^1}{\sqrt{m^2-q^2}} \, , \qquad 
\Theta^2 \mapsto \frac{\slashchar{P}\Theta^1 + q\Theta^2}{\sqrt{m^2-q^2}} \, , 
\ee
together with 
\be
e \mapsto e - \frac{2i}{\sqrt{m^2-q^2}} \bar\Theta^a \left(\slashchar{P} \delta^{ab} -q\varepsilon^{ab} \right)\dot\Theta^b\, . 
\ee
This duality transformation  takes the two spinor Noether charges  $\mathcal{Q}^a$ into two new charges $\tilde{\mathcal{Q}}^a$. 
Using the mass shell constraint, we find that 
\be
\tilde{\mathcal{Q}}^1 = - \sqrt{m^2-q^2} \, \Theta^2\, , \qquad \tilde{\mathcal{Q}}^2 = - \sqrt{m^2-q^2} \, \Theta^1\, . 
\ee
We now have four spinor Noether charges.  Using (\ref{thetaPBs}) we see  immediately that 
\be\label{N1}
\left\{ \tilde{\mathcal{Q}}^{a \, \alpha},\tilde{\mathcal{Q}}^{b \, \beta}\right\}_{PB} = -i\left[\left(\slashchar{P} \delta^{ab}+ q \varepsilon^{ab} \right)C\right]^{\alpha\beta}\, .
\ee
This assumes here that the supercharges are not chiral, because otherwise a chiral projection matrices is needed on the right 
hand side. Similarly, we find that 
\be\label{N2}
\left\{\mathcal{Q}^a_\alpha, \tilde{\mathcal{Q}}^{b \, \beta}\right\}_{PB} = \sqrt{m^2-q^2}\,  \left(\sigma_1\right)^{ab} \delta_\alpha^\beta\, . 
\ee
If we continue to assume, for simplicity of presentation, that the spinor charges are not chiral, we can use  $C$ to raise the spinor index of ${\mathcal{Q}}^a$ to arrive at an $N=4$ supersymmetry algebra of the form 
\be
\left\{ \mathcal{Q}^{A \, \alpha} , \mathcal{Q}^{B \, \beta}\right\}_{PB} = -i \left[\left(\slashchar{P} \delta^{AB} + q^{AB} \right)C\right]^{\alpha\beta}\, . 
\ee
In the basis 
$\mathcal{Q}^A= \left(\mathcal{Q}^1, \, \mathcal{Q}^2, \, \tilde{\mathcal{Q}}^1, \, \tilde{\mathcal{Q}}^2\right)$,  one finds that 
\be
q^{AB} = \left(\begin{array}{cccc} 0 & q & \sqrt{m^2-q^2} & 0 \\ -q & 0 & 0 & \sqrt{m^2-q^2} \\ -\sqrt{m^2-q^2} & 0 & 0 & -q \\ 0 & -\sqrt{m^2-q^2} & q & 0 \end{array}\right)\,,
\ee
which is an antisymmetric matrix with skew eigenvalues $\pm m$, showing that the BPS bound is saturated. 

This shows that the generic $N=2$ massive superparticle action actually has $N=4$  supersymmetry, with a BPS-saturated $N=4$ supersymmetry algebra. 
The only case for which there are no hidden supersymmetries is that for which $m=|q|$. In other words, the only $N=2$ superparticle action that really
has only $N=2$ supersymmetry is the one for which the BPS bound is saturated.  

For $N>2$ the dualities that lead to hidden supersymmetries can be found by combining the $N=1$ and $N=2$ cases. For example, consider the generic $
N=4$ superparticle action 
\be
S = \int dt \! \left\{  \dot{X}\cdot P + i\bar{\Theta}^A\left(\slashchar{P}\delta^{AB} + q^{AB}\right)\dot{\Theta}^B - \frac{1}{2}e\,(P^2 + m^2)  \right\} \,, 
\ee
where $q^{AB}$ is an antisymmetric $4\times 4$ matrix with skew-eigenvalues $q_i$ $(i=1,2)$. In a skew-diagonal basis  we have
\be
\Theta^A = \left(\begin{array}{c}\Theta^{a}_1 \\ \Theta^{a}_2 \end{array}\right)\,, \qquad (a=1,2; \ i=1,2), 
\ee
where $\Theta^a_i$ are two $2$-vectors, in terms of which the $N=4$ action becomes 
\be
S = \int dt \, \left\{  \dot{X}\cdot P + i\sum_{i=1}^2 \bar{\Theta}_i^a\left(\slashchar{P}\delta^{ab} + q_i \varepsilon^{ab} \right) \dot{\Theta}_i^a - \frac{1}{2}e\,(P^2 + m^2)  \right\} \,.
\ee
Everything now essentially reduces to the $N=2$ case on each of the two sectors labelled by $i=1,2$. In other words, we now have two separate dualities, indexed by $i$: 
\be
\Theta_i^1  \mapsto \frac{\slashchar{P}\Theta_i^2 -q_i\Theta_i^1}{\sqrt{m^2 -q_i^2}} \, , \qquad 
\Theta_i^2 \mapsto \frac{\slashchar{P}\Theta_i^1 + q_i\Theta_i^2}{\sqrt{m^2 - q_i^2}} \,  \qquad (i=1,2).
\ee
Different situations will arise based on the relative magnitudes of $m^2$ and the $q_i^2$.   When $q_1\ne q_2$ we may assume that $|q_1|>|q_2|$ and the BPS bound is $m\ge |q_1|$, and if  this bound  is saturated we have a ``half-BPS superparticle'' action. When $|q_1|=|q_2|=|q|$ the BPS bound is $m\ge |q|$, and if this is saturated we have a ``BPS superparticle'' action (which is the $N=4$ massive superparticle obtainable  by reduction from one higher dimension). 

The $N=3$ case can be analysed in the same way; after putting the central charge matrix into a standard form one gets an $N=1$ and an $N=2$ ``sector''.  More generally, there are $n$ $N=2$ ``sectors''  for $N=2n$ and an additional $N=1$ ``sector''  for $N=2n+1$. 

\subsection{Other dimensions}

Our analysis has assumed the existence of Majorana spinors, and an antisymmetric charge conjugation matrix, but the results are general.  To illustrate this, 
we consider briefly the case of $D=6$ for which the minimal spinor is $SU(2)$-Majorana-Weyl. The spin group in $D=6$ is $Sl(2;\bH) \cong SU^*(4)$ and we can use a 
4-component spinor notation in which the $SU(2)$-Majorana-Weyl superspace coordinate is  $\Theta^\alpha_i$ ($i=1,2; \alpha=1,\dots,4$) and $6$-vectors are 
antisymmetric bi-spinors (and $SU(2)$ singlets), e.g. $\bX^{\alpha\beta}$ for the Minkowski space coordinates.  In this notation, the massive superparticle action with manifest $(1,0)$ $D=6$ supersymmetry  is 
\be\label{10}
S= \int\! dt\left\{ \left(\dot {\bX}^{\alpha\beta}+ i\varepsilon^{ij}\Theta^\alpha_i \dot \Theta^\beta_j\right)  \bP_{\alpha\beta} - \frac{1}{2} e\,  \left(\bP^2 + m^2\right) \right\} \, , 
\ee
where 
\be
{\bP}^2 = \bP^{\alpha\beta}\bP_{\alpha\beta} \, , \qquad \bP^{\alpha\beta}= \frac{1}{2} \varepsilon^{\alpha\beta\gamma\delta} \bP_{\gamma\delta}\, . 
\ee
There are actually two spinor supersymmetry charges:
\be\label{original}
\mathcal{Q}_\alpha^i = \bP_{\alpha\beta}\varepsilon^{ij} \Theta^\beta_j \, ,  \qquad \tilde{\mathcal{Q}}^\alpha_i = m \Theta^\alpha_i\, . 
\ee
The first is the Noether charge of the manifest $(1,0)$ supersymmetry. The second is the Noether charge of a hidden $(0,1)$ supersymmetry. 

Now we define a new spinor variable $\Theta_\alpha^i$ of the opposite chirality, and a new einbein $\tilde e$, by 
\be\label{6Dduality}
\Theta^\alpha_i = m^{-1} \bP^{\alpha\beta} \Theta_\beta^j \varepsilon_{ji} \, , \qquad  e= \tilde e -\frac{i}{2m^2} \bP^{\alpha\beta} \Theta_\alpha^i\dot\Theta_\beta^j \varepsilon_{ji} \, . 
\ee
Using the identity
\be
\bP^{\alpha\gamma}\bP_{\beta\gamma} = \frac{1}{4} \delta^\alpha_\beta\, \bP^2\, , 
\ee
one finds that the  action (\ref{10}) becomes, when written in terms of the new variables,  the $(0,1)$ massive superparticle action 
\be\label{01}
S= \int\! dt\left\{ \left(\dot {\bX}^{\alpha\beta}+ \frac{i}{2}\varepsilon^{\alpha\beta\gamma\delta} \Theta_\gamma^i \dot \Theta_\delta^j \varepsilon_{ji}\right)  \bP_{\alpha\beta} 
- \frac{1}{2} \tilde e\,  \left(\bP^2 + m^2\right) \right\} \, , 
\ee
which has the two spinor supersymmetry charges 
\be
\mathcal{Q}^\alpha_i = \bP^{\alpha\beta} \Theta_\beta^j \varepsilon_{ji}\, , \qquad \tilde{\mathcal{Q}}_\alpha^i = m \Theta_\alpha^i \, .  
\ee
The first is the Noether charge of the manifest $(0,1)$ supersymmetry. The second is the Noether charge of a hidden $(1,0)$ supersymmetry. 
We see that the $(1,0)$ action is equivalent  to the $(0,1)$ action, and both have $(1,1)$ supersymmetry. The duality transformation (\ref{6Dduality})
that takes  the $(1,0)$ action into the $(0,1)$ action takes the manifest to the hidden supersymmetry,  and {\it vice versa}.

\section{Equivalence}

We have seen that the generic $N=n$ massive superparticle actually has $N=2n$ supersymmetrry, and that the full supersymmetry algebra is the BPS-saturated
$N=2n$ supersymmetry algebra, realised manifestly by the massive BPS $N=2n$ superparticle action. Because of the kappa-symmetry gauge invariance of the 
latter case, the (graded) dimension of the physical phase space of the two superparticle actions is also the same, which suggests a possible equivalence. For $D=3,4$ this equivalence has been proved using twistor methods \cite{Gorbunov:1996ed,Mezincescu:2013nta}. We shall now show how the  equivalence can be proved in any dimension although, 
for simplicity, we continue to assume the spinors and spinor properties of $D=3,4,10$ mod $8$. 

Like the hidden supersymmetries, the equivalence for general $N$ can be built up from the $N = 1, 2$ cases. Let us start by considering the action for the BPS-saturated $N=2$ superparticle with $q=m$:
\be\label{N=2}
S= \int \! dt \left\{ \dot X\cdot P + i \bar\Theta^a \left(\slashchar{P} \delta^{ab} + m\varepsilon^{ab}\right) \dot\Theta^b - \frac{1}{2} e\left(P^2+m^2\right) \right\}\, .
\ee
This action has manifest $N=2$ supersymmetry but a non-manifest kappa-symmetry gauge invariance. The infinitesimal gauge transformations, with 
arbitrary spinor parameter $\kappa(t)$, are 
\be 
\delta_{\kappa} X = -i\bar{\Theta}^a\Gamma^{\mu}\delta_\kappa \Theta^a\,, \qquad \delta_{\kappa} \Theta^a = \left(\slashchar{P}\delta^{ab} - m\varepsilon^{ab}\right)\kappa^b \,, \qquad 
\delta_{\kappa} e = -4i\bar{\kappa}^a\dot{\Theta}^a \,.
\ee
Notice that, on the mass-shell, 
\be
\delta_\kappa \Theta^1 = \hat\kappa \, , \quad \delta_\kappa\Theta^2 = - m^{-1}  \slashchar{P}\hat\kappa\, \qquad \hat\kappa \equiv 
\slashchar{P}\kappa^1 -m\kappa^2\, ,
\ee
which shows that the two spinor parameters $(\kappa^1,\kappa^2)$ occur only through the one linear combination $\hat\kappa$.  Notice also that 
$\delta_{\kappa} \left(\slashchar{P}\Theta^1 + m\Theta^2\right) =0$, which  suggests that we may fix the gauge by the condition
\be
\slashchar{P}\Theta^1 - m\Theta^2 = 0 \,. 
\ee
This is indeed a valid gauge condition since the kappa-symmetry gauge transformation of the left hand side is $2\slashchar{P} \hat\kappa$. If we use 
this gauge condition to eliminate $\Theta^2$, and define
\be
\Theta = 2\Theta^1 \, , \qquad  e'= e+ \frac{2i}{m^2}\bar\Theta^1 \slashchar{P}\Theta^1\, , 
\ee
then we find that  the action (\ref{N=2}) reduces to 
\be
S= \int \! dt \left\{\left(\dot X + i\bar\Theta \Gamma\dot\Theta\right) \cdot P  - \frac{1}{2} e'\left(P^2+m^2\right) \right\}\, , 
\ee
which is action of the massive $N = 1$ superparticle. 
The supercharges of the $N = 2$ action reduce to the two supercharges, manifest and hidden, of this $N=1$  action
\bea
\mathcal{Q}^1 &=& \slashchar{P}\Theta^1 + m \Theta^2 = \slashchar{P}\Theta = \mathcal{Q} \,, \nonumber \\
\mathcal{Q}^2 &=& \slashchar{P}\Theta^2 - m \Theta^1 = -m\Theta = \tilde{\mathcal{Q}} \,.
\eea

To illustrate how this works for higher $N$, we shall consider the massive $N=4$ superparticle. Let $\left\{ q_i;i=1,2\right\}$ be the skew-eigenvalues of the central charge matrix. 
In the skew-eigenbasis, the action is 
\be\label{N=4}
S= \int \! dt \left\{ \dot X\cdot P + i \sum_{i=1}^2\bar{\Theta}_i^a \left(\slashchar{P} \delta^{ab} + q_i\varepsilon^{ab}\right) \dot\Theta_i^b - \frac{1}{2} e\left(P^2+m^2\right)  \right\}\, , 
\ee
where we may assume  (given unitarity) that  $q_1\ge q_2 \ge 0$. There are then three possibilities:
\begin{itemize}

\item If $q_1=q_2=  m$, the action (\ref{N=4}) is invariant under the kappa-symmetry gauge transformation
\be
\delta_{\kappa} X = -i\sum_{i=1}^2\bar{\Theta}_i^a\Gamma^{\mu}\delta_{\kappa_i} \Theta_i^a\,, \qquad 
\delta_{\kappa} e = -4i\sum_{i=1}^2 \bar{\kappa}_i^a\dot{\Theta}_i^a \,,
\ee
where 
\be
\delta_{\kappa_i} \Theta_i^a = \left(\slashchar{P}\delta^{ab} - m\varepsilon^{ab}\right)\kappa_i^b \qquad (i=1,2).
\ee
There are now two pairs of spinor parameters,  $\kappa_i^a$ and $\kappa_2^a$, and each of the corresponding two gauge invariances  may be 
fixed separately, in the manner described above. Fixing both leads to the massive $N=2$ superparticle action with no Wess-Zumino mass term:
\be
S= \int \! dt \left\{ \left(\dot X + i \bar\Theta^a\Gamma \dot\Theta^a\right)\cdot P  - \frac{1}{2} e\left(P^2+m^2\right) \right\}\, .
\ee
Alternatively, one could fix only one of the two kappa-symmetries, leaving the massive $N=3$ ``partially BPS-saturated'' superparticle action
\be\label{N=3BPS}
S= \int \! dt \left\{ \dot X\cdot P + i \bar\Theta_1 \slashchar{P} \dot{\Theta}_1  + i \bar\Theta_2^a \left(\slashchar{P} \delta^{ab} 
+ m\varepsilon^{ab}\right) \dot\Theta_2^b - \frac{1}{2} e\left(P^2+m^2\right) \right\}\, .
\ee

\item  If $q_1= m$ but $q_2 =q < m$, the action (\ref{N=4}) has only one kappa-symmetry. After gauge fixing it reduces to the $N=3$ non-BPS action
\be\label{N=3}
S= \int \! dt \left\{ \dot X\cdot P + i \bar\Theta_1 \slashchar{P} \dot{\Theta}_1  + i \bar\Theta_2^a \left(\slashchar{P} \delta^{ab} 
+ q\varepsilon^{ab}\right) \dot\Theta_2^b - \frac{1}{2} e\left(P^2+m^2\right) \right\}\, .
\ee
This shows that the generic non-BPS $N=3$ superparticle action is equivalent to the  partially-BPS $N=4$ superparticle action, but both actually have $N=6$ supersymmetry and are equivalent, via gauge-fixing, to the $N=6$ BPS superparticle. 

\item Finally, If both  $q_1< m$ and $q_2 < m$, the  $N=4$ action is non-BPS and has no  kappa-symmetry to fix. This action is equivalent to the $N=8$ BPS superparticle action. 
The proof of equivalence in this case is a straightforward extension of the proof of equivalence of the $N=1$ and $N=2$ BPS actions.

\end{itemize}
The general pattern should be clear. The $N = 2n$ BPS superparticle has  $n$ kappa-symmetries (of the $N=2$ type), any of which may be fixed to get an action with a fewer number of manifest supersymmetries. In this way, all actions for massive non-BPS superparticles for which the total number of supersymmetries is $2n$ may be shown to be equivalent to the $
N=2n$ BPS superparticle action. This is true for any $n$ so we conclude that  {\it all (unitary) massive superparticles are either BPS  or else gauge-fixed, or partially gauge-fixed, versions of a BPS massive superparticle, for  which all supersymmetries are manifest}.

\section{Discussion}

The massless superparticle action, which is that of (\ref{N=1}) with $m=0$,   provides a classical description of a supermultiplet of massless 
particles in a D-dimensional spacetime, here taken to be 
Minkowski spacetime. The fact that massless supermultiplets are ``short'' relative to massive supermultiplets is a consequence of a fermionic gauge invariance 
(called kappa-symmetry) of the superparticle action.  Dimensional reduction (achieved by fixing some components of the momentum) yields a lower-dimensional 
massive superparticle  action with similar properties: the extra momentum components appear as central charges in the lower-dimensional supersymmetry 
algebra, which is ``BPS-saturated''; i.e. the mass is given in terms of the central charges so as to saturate a unitarity bound implied by the supersymmetry algebra.
Quantization then yields a ``short'' (or ``BPS'') supermultiplet, again as a consequence of a fermionic gauge invariance of the classical superparticle action.

In contrast, generic massive superparticle actions do not have a fermionic gauge invariance, and their quantization yields a ``long'' supermultiplet. Nevertheless,  
we have shown here that the distinction betwen a ``BPS superparticle'' (with a kappa-symmetric action) and a ``non-BPS superparticle'' is illusory. Essentially, this 
is because any ``long'' supermultiplet is also a ``short'' supermultiplet of some other supersymmetry algebra with additional supersymmetries. This fact gets realized
at the level of classical particle mechanics through ``hidden'', i.e. non-manifest, supersymmetries that are a consequence of dualities of non-BPS superparticle actions. 
These dualities are either discrete symmetries or they exchange one action with chiral spinor variables for an equivalent one with anti-chiral spinor variables; in either 
case the manifest  supersymmetry Noether charges are exchanged with the  ``hidden'' supersymmetry Noether charges. 

A simple example is the massive $N=1$ superparticle for $D=10$. In principle there are two such actions, according to the chirality of the  spinor supersymmetry charge; i.e. according to whether we interpret $N=1$ as $(1,0)$ or as $(0,1)$. In fact, the $(1,0)$ and $(0,1)$ actions are equivalent (in the sense of being related by field redefinitions) because they are exchanged by a massive superparticle duality, and this unique $N=1$ massive superparticle action actually has $(1,1)$ supersymmetry.  This means that its 
full supersymmetry algebra is the same as that of the kappa-symmetric $(1,1)$ BPS-superparticle that governs the dynamics of an isolated D0-brane of IIA superstring theory.
And because of the latter's kappa-symmetry gauge invariance, the two superparticles also have a physical phase space of the same (graded) dimension. 
In fact, it has been shown previously for  $D=3$ and $D=4$, using twistor methods applicable in these dimensions,  that the $N=1$ massive superparticle is equivalent to the $N=2$ BPS massive superparticle. 

An important caveat here is that  {\it quantum}  equivalence  of the massive $N=1$ and BPS-saturated $N=2$ superparticles applies only if one allows, in both cases,  for the same ambiguities in passing from  the classical to the quantum theory. The short supermultiplet of $N=2$ supersymmetry allowed when $m=|q|$ carries a non-zero central charge. This means that if the superparticle is quantized as an $N=2$ theory then its wavefunction must be complex, which leads to a doubling of the states, which can be separated into positive and negative eigenstates of the central charge. If it is quantized as an $N=1$ theory then the wavefunction can be chosen to be real, in which case it describes an $N=1$ supermultiplet of zero  superspin; in this case the hidden $N=2$ supersymmetry of the classical $N=1$ superparticle is lost when passing to the quantum theory. There is really only one classical theory but there are two inequivalent ways of quantizing it. 

Here we have shown that  the  equivalence of the $N=1$ massive superparticle  to the $N=2$ massive BPS superparticle is a general result, valid in {\it any} space-time dimension; assuming (in the case of chiral supersymmetry algebras) that $N=2$ means $(1,1)$ when $N=1$ means $(1,0)$ or $(0,1)$. The former is just a gauge-fixed version of the latter. 
Moreover, this  is just a particular case of a much more general equivalence between any two superparticle actions with the same full supersymmetry algebra; all non-BPS massive superparticles are gauge-fixed or partial gauge fixed  BPS massive superparticles.  If we consider the massless superparticle as trivially ``BPS'' then we arrive at the conclusion that ``all superparticles are BPS''.

\section*{Acknowledgements}

We are grateful to  Joaquim Gomis and Michael Green for useful discussions.  L.M. acknowledges partial support from National Science Foundation Award  PHY-1214521. 
A.J.R. acknowledges support from the UK Science and Technology Facilities Council.


\providecommand{\href}[2]{#2}\begingroup\raggedright
\endgroup

\end{document}